\documentclass{DISproc}
\def \et {E_{T}}

\def  \met {\not\!\!\et }

\begin{document}
%------------------------------------

\title{Search for the Standard Model Higgs Boson in ATLAS}
%for single authors the superscripts are optional
\author{{\slshape Prolay Mal}\\[1ex]
(on behalf of the ATLAS collaboration)\\
SPP/IRFU, CEA Saclay, Gif-sur-Yvette 91191, France}
% please enter the contribution ID for the DOI
\contribID{75}
\doi  % if there is an online version we will register DOIs
\maketitle
\begin{abstract}
The search for the Standard Model (SM) Higgs boson
based on 4.7-4.9 $\rm fb^{-1}$ of pp collision data at
$\rm\sqrt{s}=7$ TeV recorded with the ATLAS detector
is presented here. The combined ATLAS results exclude the
SM Higgs boson masses ($\rm m_H$) of 110.0-117.5, 118.5-122.5
and 129-539 GeV at 95\% confidence level. An excess of
events has been observed around $\rm m_H\sim 126$ GeV with
the global probability of 30\% (10\%) to occur anywhere in
$\rm 110<m_H<600$ ($\rm 110<m_H<146$) GeV.
\end{abstract}

\section{Introduction}
The Higgs mechanism~\cite{higgsmech} provides a general framework
to explain the observed masses of the W and Z gauge
bosons through electroweak symmetry breaking. Within the Standard
Model (SM), this mechanism posits the existence of a scalar
boson, the Higgs boson, with an a priori unknown mass ($\rm m_H$).
The direct searches for the SM Higgs boson at the LEP experiments
have excluded  $\rm m_H<114.4$ GeV~\cite{lepwg}, while the searches
at the Tevatron exclude $\rm 156<m_H<177$ GeV~\cite{tevwg}.
However, a global fit of the electroweak measurements performed at LEP,
SLD and the Tevatron experiments, predicts a SM Higgs boson mass of
$\rm 94^{+29}_{-24}$ GeV.
\section{Individual Search Channels}
The SM Higgs searches in ATLAS have been performed over a wide range of
Higgs boson masses (110-600 GeV) considering different SM production
mechanisms (gluon fusion, vector boson fusion and vector boson associated
production) and their subsequent decay modes. The detector resolution for the
reconstructed Higgs boson mass plays a crucial role in classifying
the searches into numerous channels having different selection criteria
as detailed below. The SM Higgs boson signal events have
been simulated using PowHeg and Pythia generators, while the background
contributions have been estimated either using simulation, or directly
from data as appropriate.
\subsection{$H\rightarrow ZZ\rightarrow l^+l^-\nu\bar{\nu}$,
$H\rightarrow ZZ\rightarrow l^+l^-q\bar{q}$,
$H\rightarrow WW\rightarrow l\nu q\bar{q}^\prime$}
Searches in these channels are focused on high mass Higgs boson searches
over a typical Higgs mass range of 200-600 GeV. In the ZZ channels the
events are required to have a lepton pair ($\rm e^+e^-$ or $\rm\mu^+\mu^-$)
with reconstructed $\rm M_{ll}$ close to the Z boson mass. The selected
events for the $\rm H\rightarrow ZZ\rightarrow l^+l^-\nu\bar{\nu}$
\cite{zzllnunu} search are further classified into two subcategories
considering the pile-up effects (`low' and `high') on the reconstructed
missing transverse energies ($\rm\met$), while the 
$\rm H\rightarrow ZZ\rightarrow l^+l^-q\bar{q}$~\cite{zzllqq}
search considers `tagged' (2 b-tagged jets) and `untagged'
($\rm < 2$ b-tagged jets) events separately.
Figures~\ref{Fig:searchchannels} (a) and~\ref{Fig:searchchannels} (b) show
the transverse mass distribution for
$\rm H\rightarrow ZZ\rightarrow l^+l^-\nu\bar{\nu}$ candidates in the
`low' pile-up data, and $\rm m_{l^+l^-q\bar{q}}$ distribution for
$\rm H\rightarrow ZZ\rightarrow l^+l^-q\bar{q}$ in `tagged' selection
respectively. The selection criteria for the
$\rm H\rightarrow WW\rightarrow l\nu q\bar{q^\prime}$ search~\cite{wwlnuqq}
are optimized over a Higgs boson mass range of 300-600 GeV considering 
`H+0jet', `H+1jet' and `H+2jets' (mostly from the vector boson fusion
processes).
\subsection{$H\rightarrow ZZ^{(*)}\rightarrow l^+l^- l^{\prime +}l^{\prime -}$}
The search in this channel~\cite{zz4l} consists of event categories
with different lepton flavor combinations, while the SM $ZZ^{(*)}$ production
processes remain irreducible background at the final level of event selection.
Full reconstruction of the Higgs boson mass is possible for this channel with
excellent mass resolution (2\% and 1.5\% for 4e and 4$\mu$ at
$m_H\sim120$ GeV). The 4-lepton invariant mass ($m_{4l}$) distribution is
displayed in Figure~\ref{Fig:searchchannels} (c).
\subsection{$H\rightarrow \gamma\gamma$}
\begin{figure}[htb]
  \centering
\begin{tabular}{ccc}
\includegraphics[width=1.75in,height=1.6in]{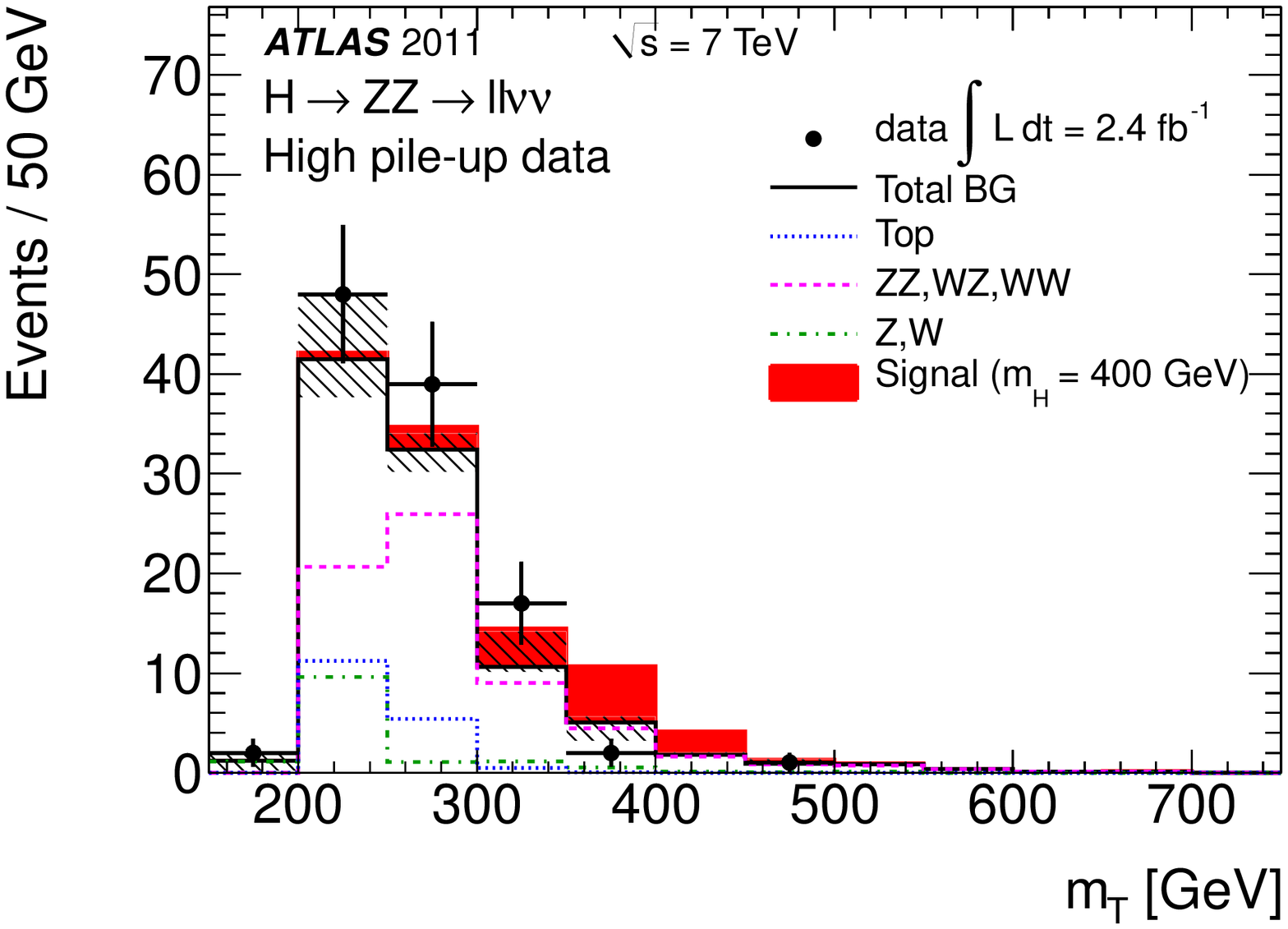}
&\includegraphics[width=1.75in,height=1.6in]{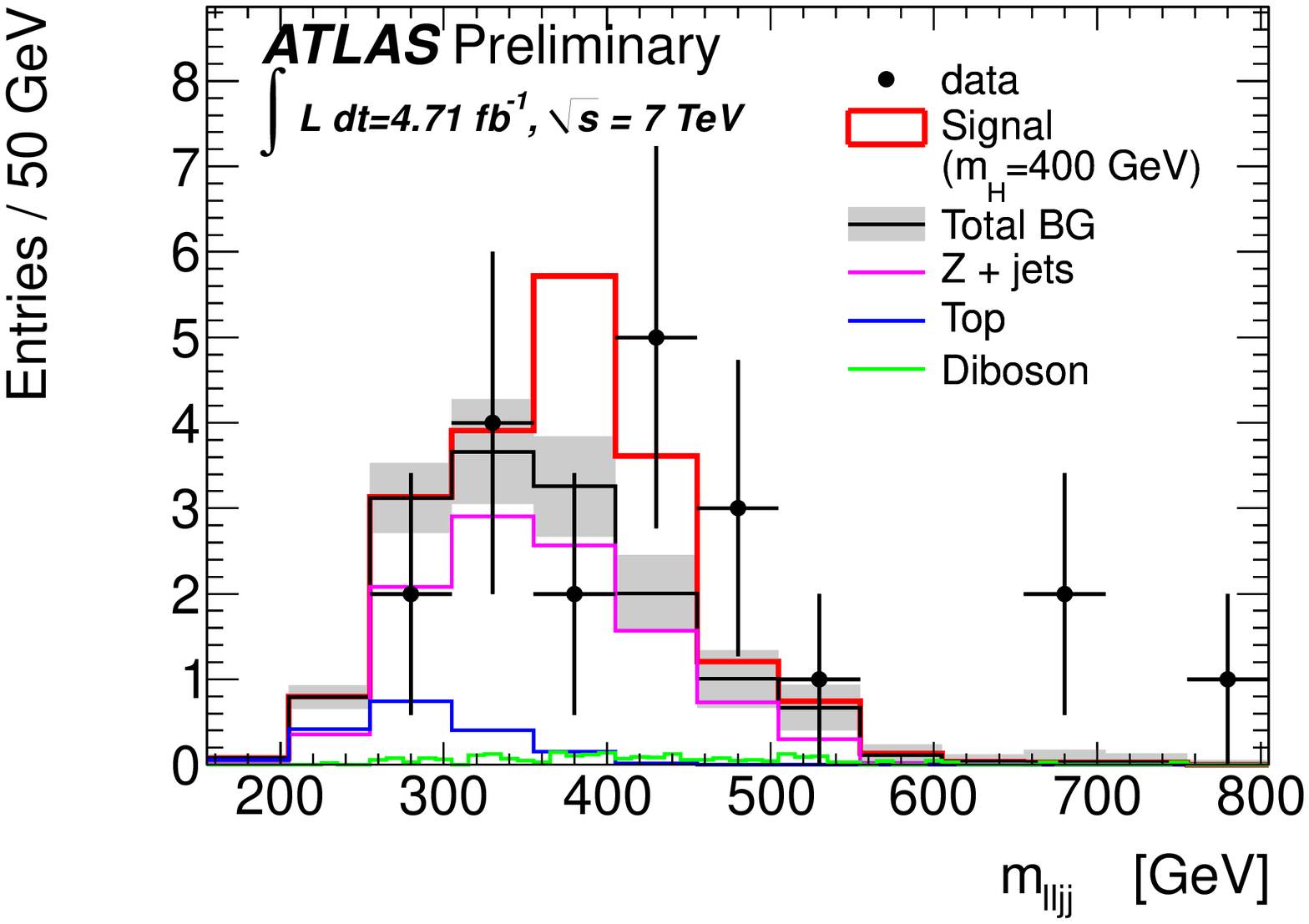}
&\includegraphics[width=1.75in,height=1.6in]{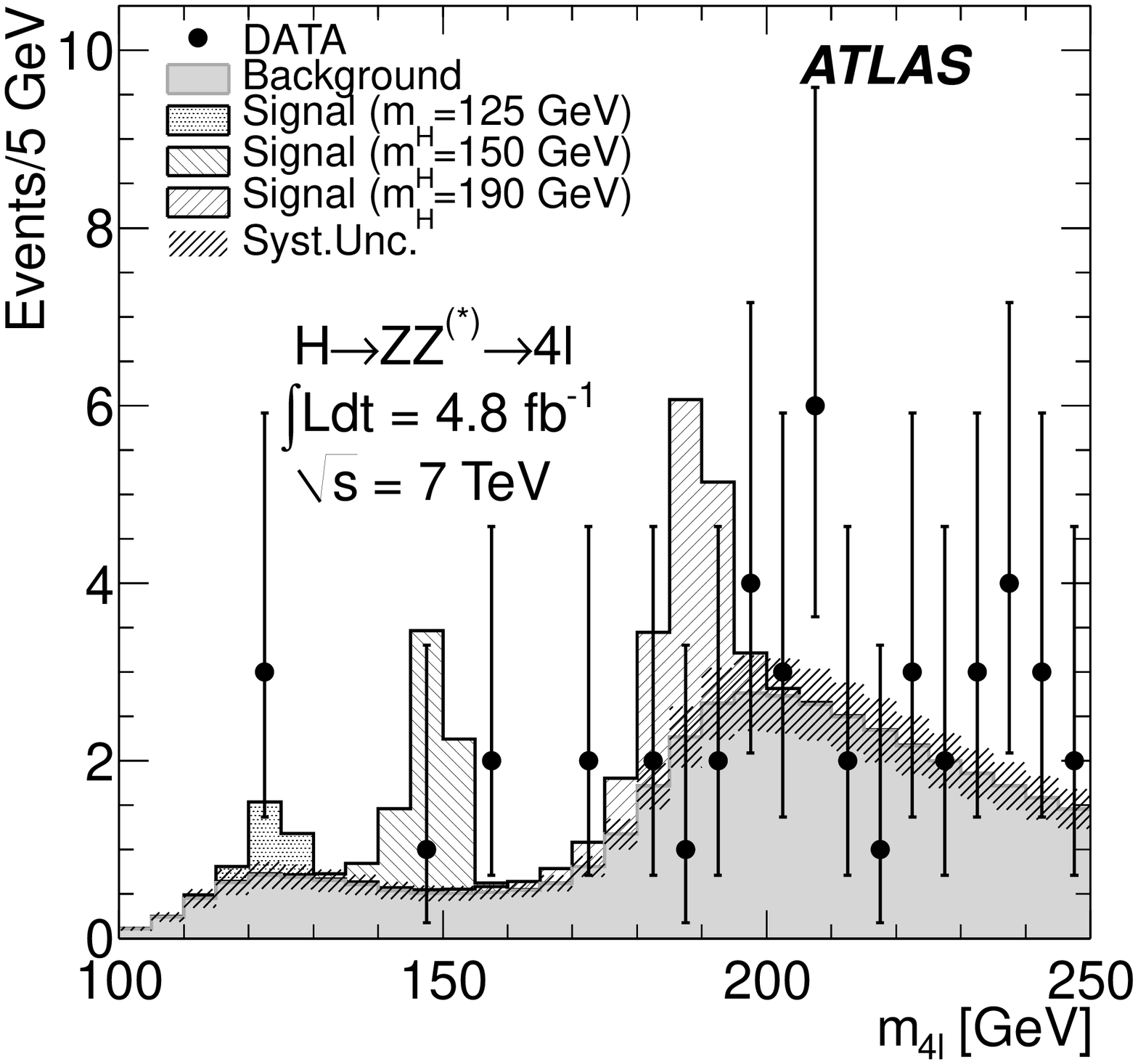}\\
(a) & (b) & (c) \\
\includegraphics[width=1.75in,height=1.5in]{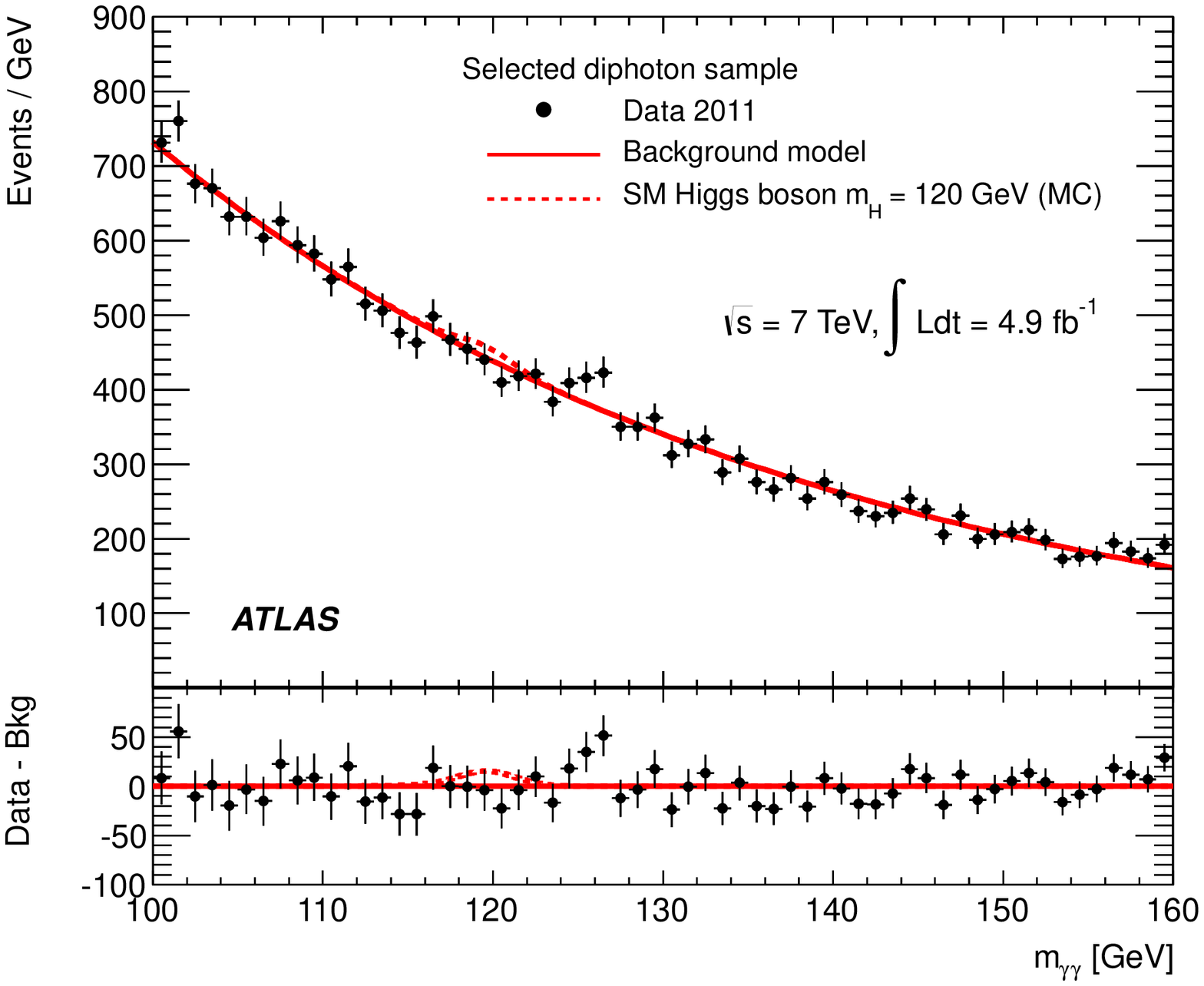}
&\includegraphics[width=1.75in,height=1.65in]{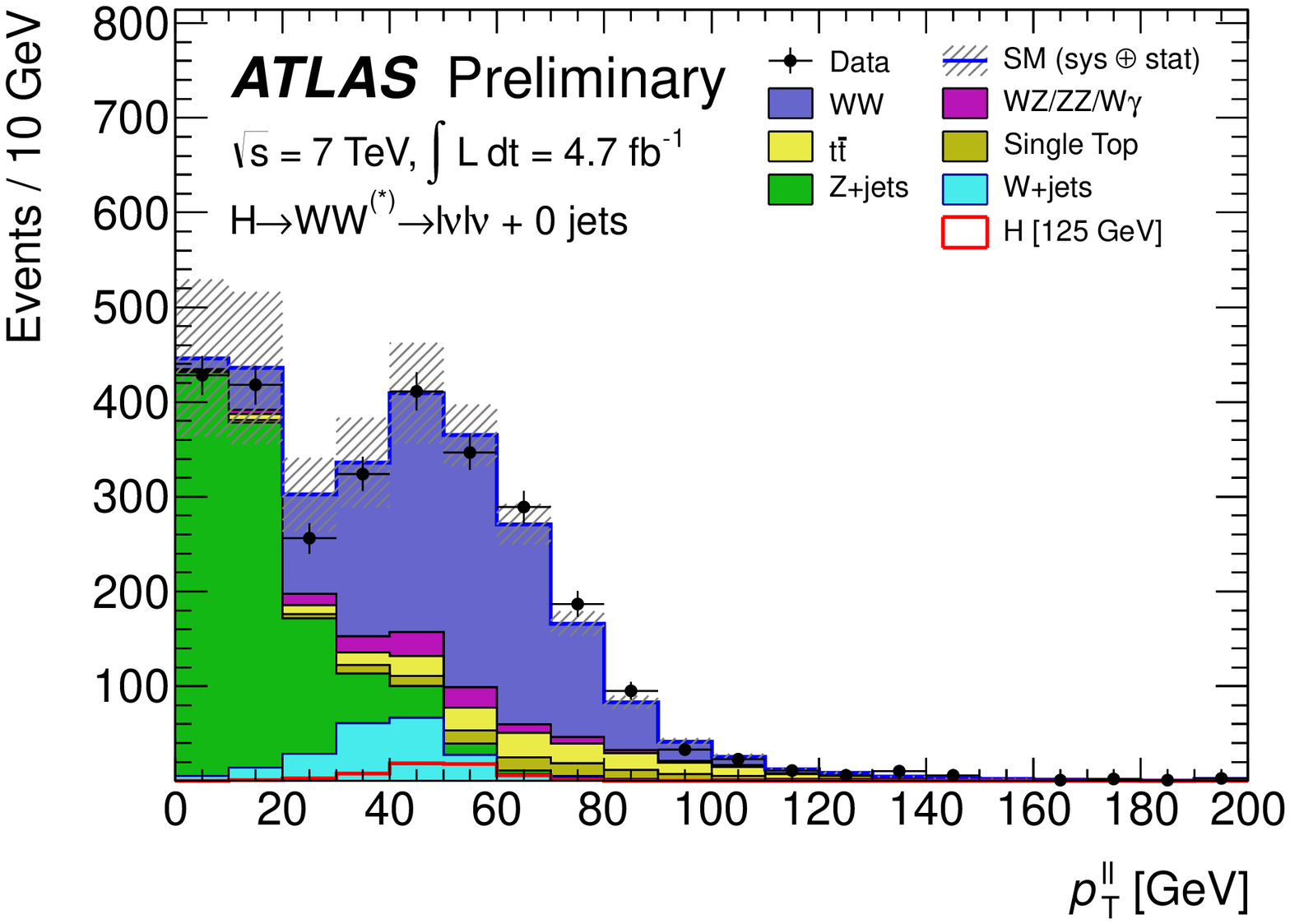}
&\includegraphics[width=1.75in,height=1.6in]{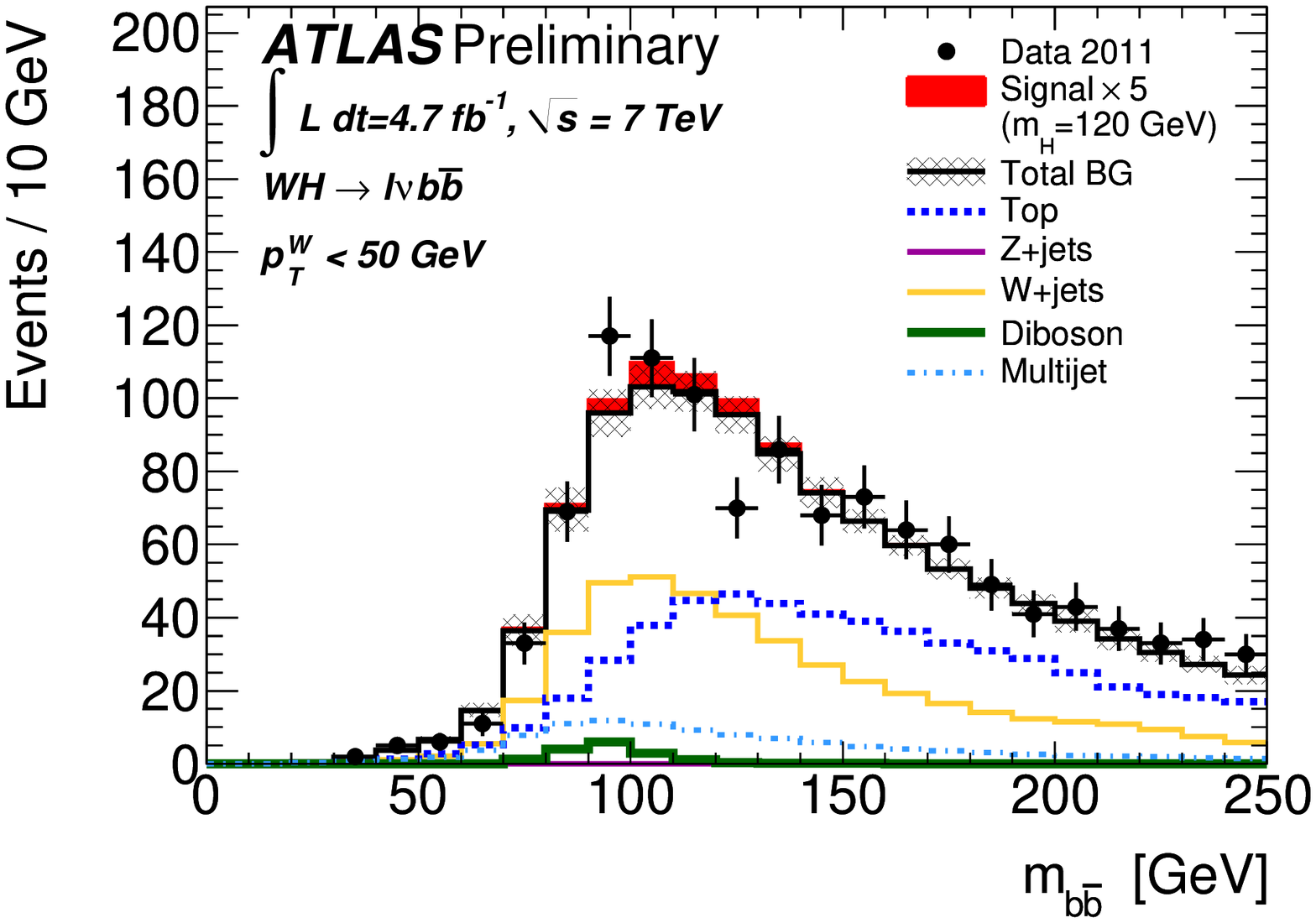}\\
(d) & (e) & (f) \\
\end{tabular}
  \caption{ATLAS SM Higgs Search observables in different channels:
(a) the transverse mass distribution in
$H\rightarrow ZZ\rightarrow l^+l^-\nu\bar{\nu}$ channel, (b) the invariant
mass distribution of the llqq system for the
$H\rightarrow ZZ\rightarrow l^+l^-q\bar{q}$ search, (c) the $\rm m_{4l}$
distribution for the
$H\rightarrow ZZ^{(*)}\rightarrow l^+l^- l^{\prime +}l^{\prime -}$ candidate
events, (d)the $\rm m_{\gamma\gamma}$ spectrum in $H\rightarrow\gamma\gamma$
search, (e) the $\rm m_T$ distribution in the
$H\rightarrow WW^{(*)}\rightarrow l^+\nu l^{\prime -}\bar{\nu}$, and (f) the
$\rm m_{b\bar{b}}$ distribution for the $WH\rightarrow l\nu b\bar{b}$
analysis.}
  \label{Fig:searchchannels}
\end{figure}

Although $\rm H\rightarrow \gamma\gamma$ decays have small branching
ratio (about 0.2\%), search in this channel~\cite{gammagamma} has the
potential to discover the Higgs boson in the low mass range (110-150 GeV).
The analysis is split into nine independent sub-channels based on the
photon pseudorapidity, conversion status, and the momentum component of
the diphoton system transverse to the thrust axis ($\rm p_{Tt}$). The
background distribution is obtained by fitting the $\rm m_{\gamma\gamma}$
distribution in data with a smoothly falling exponential function, while
the ATLAS $\rm m_{\gamma\gamma}$ mass resolution is approximately
1.4\% for $\rm m_H=120$ GeV. The diphoton invariant mass
distribution is shown in Figure~\ref{Fig:searchchannels} (d).
\subsection{$H\rightarrow WW^{(*)}\rightarrow l^+\nu l^{\prime -}\bar{\nu}$}
The searches in this channel~\cite{wwlnulnu} cover a wide range
of $\rm 110<m_H<600$ GeV consisting of different number of jets
(0, 1 and 2 jets) and lepton flavor combinations (ee, $\rm\mu\mu$
and $\rm e\mu$) in the final states. The reconstructed WW transverse mass
($\rm m_T$) has been utilized in this analysis as a discriminating
variable as shown in Figure~\ref{Fig:searchchannels}(e) for the `H+0jet'
sub-channel.
\subsection{$(W/Z) H\rightarrow (l\nu/l^+l^-, \nu\bar{\nu}) b\bar{b}$}
The searches in $\rm (W/Z) H\rightarrow (l\nu/l^+l^-, \nu\bar{\nu}) b\bar{b}$
channels~\cite{vhbbbar} are optimized for $\rm m_H$ of 110-130 GeV. The
main advantage with these channels is due to the large
$\rm BR(H\rightarrow b\bar{b})$ at low $\rm m_H$, along with the
possibility to fully reconstruct the Higgs boson mass through
$\rm m_{b\bar{b}}$. The final states with one or two charged leptons are
sub-categorized depending on the transverse momentum of the reconstructed
vector boson, and the lepton flavors. Figure~\ref{Fig:searchchannels} (f)
shows the $\rm m_{b\bar{b}}$ distribution for the
$\rm WH\rightarrow l\nu b\bar{b}$ sub-channel.
\subsection{$H\rightarrow \tau^+\tau^-\rightarrow l^+l^-\nu\nu\nu\nu$,
$l\tau_{had}\nu\nu\nu$, $\tau_{had}\tau_{had}\nu\nu$}
$\rm H\rightarrow \tau^+\tau^-$ searches~\cite{tautau} consist of all possible
leptonic and hadronic ($\rm\tau_{had}$) decay modes of the $\rm\tau$-leptons
originating from the Higgs boson decay. For the
$\rm H\rightarrow \tau^+\tau^-\rightarrow l^+l^-\nu\nu\nu\nu$ and
$\rm H\rightarrow \tau_{had}\tau_{had}\nu\nu$ channels, invariant mass of
the $\rm\tau^+\tau^-$ system is reconstructed assuming collinear
approximation. The searches in $\rm H\rightarrow l\tau_{had}\nu\nu\nu$ channel
reconstruct the $\rm m_{\tau^+\tau^-}$ using the Missing Mass Calculator
techniques~\cite{mmc} where the full event topology is reconstructed using the
kinematics of the $\rm\tau$-lepton decay products.
\section{Exclusion limits}
\begin{figure}[htb]
\centering
\includegraphics[width=2in,height=1.5in]{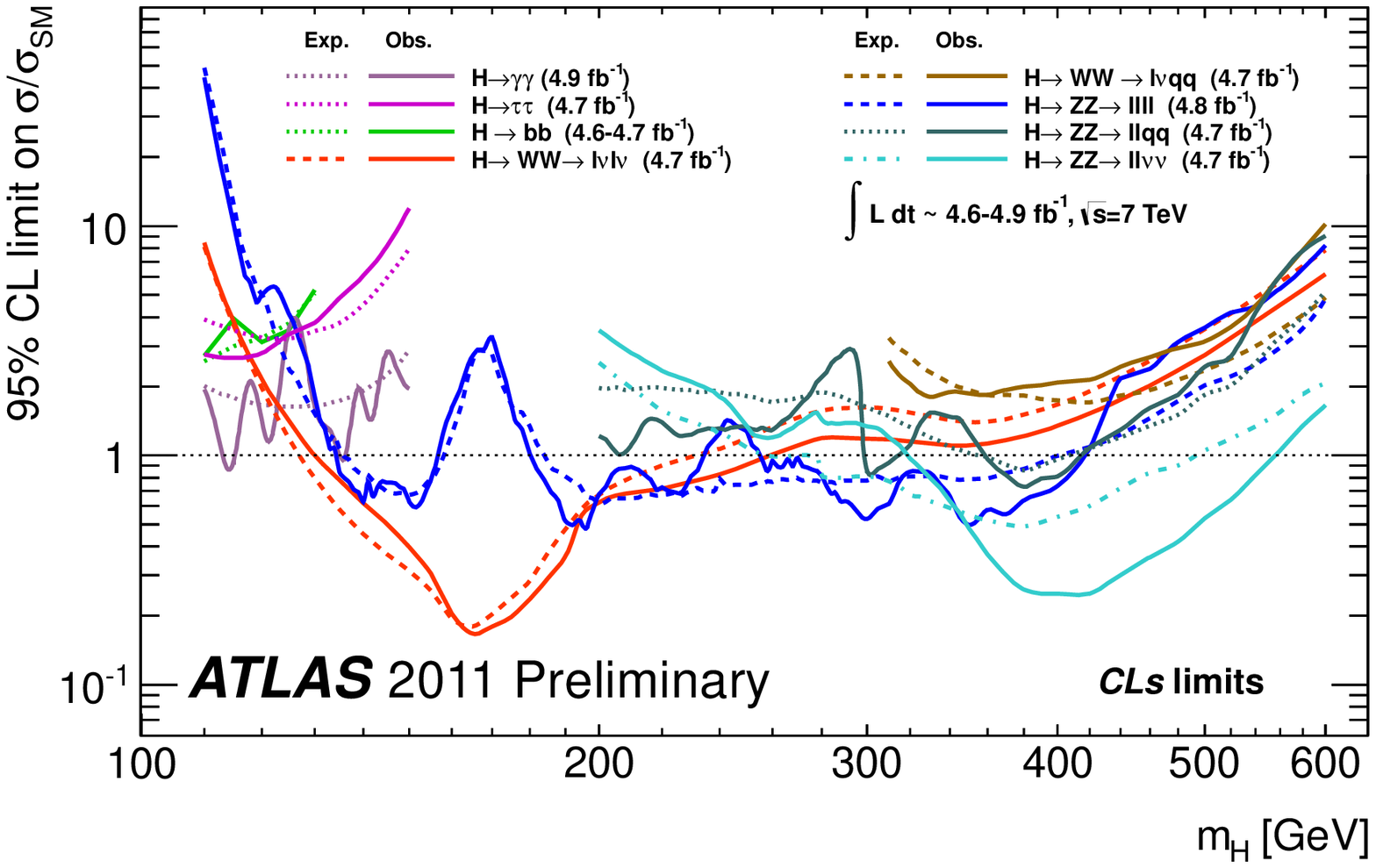}
\includegraphics[width=2in,height=1.5in]{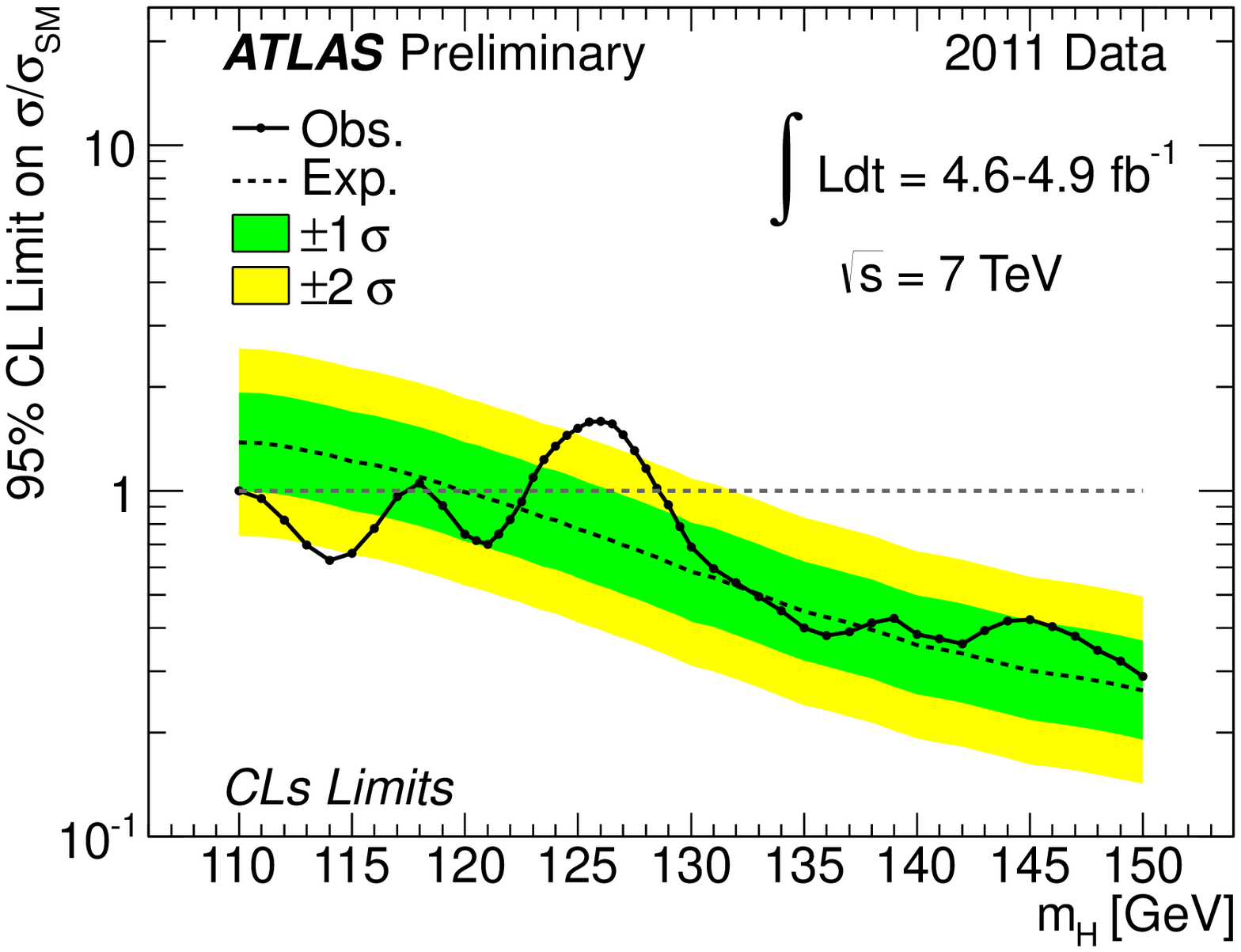}
\caption{Exclusion limits on the SM Higgs production cross-sections at 95\% CL
for individual search channels (left), and for the combination of all search
channels in $\rm m_H$ range of 110-150 GeV (right).}
\label{Fig:combination}
\end{figure}

The results from the aforesaid search channels have been utilized to set
an upper limit on the SM Higgs boson production cross section as a function
of the $\rm m_H$. The limits are conveniently expressed in terms of the
signal strength ($\rm\mu=\sigma/\sigma_{SM}$), the ratio of a given Higgs
boson production cross section ($\rm\sigma$) to its SM predicted value
($\rm\sigma_{SM}$). The $\rm CL_s$ prescription\cite{clsprescription} with
a profile likelihood ratio test statistic, $\lambda(\mu)$~\cite{profilelhood}
has been utilized here to derive the exclusion limits.
Figure~\ref{Fig:combination} shows the exclusion limits on $\mu$ at 95\%
confidence level (CL) for individual search channels, along with the
combined ones from all search channels over the $\rm m_H$ range of 110-150
GeV. A small excess of events near $\rm m_H\sim 126$ GeV is observed in
$H\rightarrow \gamma\gamma$ and
$H\rightarrow ZZ^{(*)}\rightarrow l^+l^- l^{\prime +}l^{\prime -}$ search
channels, both of which fully reconstruct the Higgs boson mass with high
resolution.

\section{Conclusions}
ATLAS has performed extensive searches for the SM Higgs boson utilizing
the full 4.7-4.9 $\rm fb^{-1}$ dataset recorded during the 2011 LHC operations.
The ATLAS combination~\cite{smhiggscomb} from numerous search channels excludes
the SM Higgs boson mass in ranges of 110.0-117.5 GeV, 118.5-122.5 GeV, and
129-539 GeV at the 95\% CL, while an exclusion of 120 GeV$<m_H<$555 GeV
is expected in the absence of Higgs signal. Furthermore, the exclusion limits
have been recalculated at 99\% CL and the SM Higgs boson over a mass range
of 130-486 GeV has been excluded. 
The excess in the observed data has a local significance of 2.5$\sigma$,
where the expected significance in the presence of a SM Higgs boson with
$\rm m_H=126$ GeV is 2.9$\rm\sigma$. The global probability for such an
excess to occur across the entire SM Higgs mass range (110-600 GeV) is
estimated to be 30\%. However, the said probability reduces to nearly
10\% if a 110-146 GeV mass range for the SM Higgs boson is considered.

{\begin{footnotesize}
% IF YOU DO NOT USE BIBTEX, USE THE FOLLOWING SAMPLE SCHEME FOR THE REFERENCES
% ----------------------------------------------------------------------------

% ----------------------------------------------------------------------------

% IF YOU USE BIBTEX,
% - DELETE THE TEXT BETWEEN THE TWO ABOVE DASHED LINES
% - UNCOMMENT THE NEXT TWO LINES AND REPLACE 'smith_joe.bib' WITH YOUR
%   FILE(S)

% \bibliographystyle{DISproc}
% \bibliography{smith_joe.bib}
\end{footnotesize}
}

% ****************************************************************************
% END OF BIBLIOGRAPHY AREA
% ****************************************************************************

\end{document}